# An Algorithm for a Variation of the Shortest Common Superstring Problem

Arthur Gilfanov
Boulder, Colorado

**Abstract:** This study develops an algorithm to solve a variation of the Shortest Common Superstring (SCS) problem. There are two modifications to the base SCS problem. First, one string in the set *S* is allowed to have up to *K* mistakes, defined as not matching the SCS in at most *K* positions. Second, no string in *S* can be a substring of another in *S.* The algorithm proposed for the problem is exact.

## 1. Introduction:

This paper will be structured in the following manner. Section 1 is the introduction. In Section 2 the problem will be technically defined. Section 3 will be the preliminaries section which will similarly define variables and other important information used throughout the study. Section 4 will present the algorithm itself, broken down into its subsections. Within this part, the time complexities of the algorithm will be described and proved. The standard SCS problem has approximation algorithms[1], and the base SCS problem[2-5] is utilized in bioinformatics, primarily in applications related to DNA sequencing and similar[6-8] and data compression[9-10]. The goal of this study is to develop an exact solution to a specific modification of the SCS problem, call it the $SCS_K$ problem. It will be defined in depth in the following section.

## 2. The Problem:

Given *n* strings in the input set S: $\{s_0, s_1, \dots, s_{n-1}\}$ find the shortest string *L* such that all but one string from *S* are an exact substring of *L*. The remaining string from *S* must fit into *L* with a maximum allowed of *K* mistakes. A mistake is a place where two strings do not match: in more detail, given two strings *a* and *b*, if they are set in a position such that index *i* of *a* coincides with index *j* of *b*, then $a_i \neq b_j$. *K* has no predetermined limit. There is one final constraint on the problem: no string in *S* can be a substring of another in *S*. Define this as the $SCS_K$ problem, which is a variation of the classical SCS problem.

**3. Preliminaries:**

Before the algorithm itself is presented, some of the essential variables and notation of the paper will be defined, along with other technical knowledge important to the study. Other information that is not stated here will be defined during the explanation of the algorithm itself. As stated above, *S* is the set of input strings: $\{s_0, s_1, \dots, s_{n-1}\}$. Given a string *a*, $a_i$ is defined as the character at position *i* in a, using 0-based indexing. $l_i$ is defined as the length of the *i*th element in *S*, again using 0-based indexing. *M* is defined as $\sum_{i=0}^{n-1} s_i$, or in other words, the sum of all string lengths in set *S,* and *c* is defined as the length of the longest string in *S.* Let $|a|$ be a notation for the length of a string, *a*, or the length of a list, *a*. The code of the algorithm will reference some other simple algorithms that are common knowledge. They will be called *MIN(a, b)*, and *BINARYSEARCH(a, x)*. *MIN(a, b)* takes two integers and returns the smaller one, and *BINARYSEARCH(a, x)* returns the first index of a value greater than x in a sorted list a, 0-based once again. If no element in *a* is greater than x, then |*a*| is returned.

**4. The Algorithm:**

There are four main sections in the algorithm.

**Part 1: Computing the amount of mistakes for each pair of strings in every position.**

Define $MISTAKES_{i,\,j,\,k}$ to be a list of the indices of mistakes (mismatches), as defined above, for strings $s_i$ and $s_j$ when they are set in a position where the last character of $s_j$ matches up with the *k th* character in $s_i$. Note: in larger values of *k* the last character of $s_j$ may not actually overlap with *k th* character in $s_i$. In this case it is taken as if there was a *k th* character since parts of the string may still overlap. In other words, Also, note that $|MISTAKES_{i,\,j,\,k}|$ can be used to calculate how many mistakes one string in *S* has compared to another string in *S* when overlaid in any position. Now, the current goal is to calculate $MISTAKES_{i,\,j,\,k}$ for every pair *i, j* and for each *k.* To do this, iterate over all pairs, and all subsequent possible overlap positions, counting the number of mistakes. Let the function that does this be called *MISTAKES_FOR_PAIRS*, implementation provided below in Algorithm Code 1. Its time complexity is stated and proved below in Lemma 1.

**Lemma 1:** *Time complexity of MISTAKES_FOR_PAIRS is* $O(n^2c^2)$. *As stated above, c is the length of the maximum length string in S.*

*Proof:* The algorithm iterates over all pairs *i, j*. This takes $O(n^2)$. Next, for each pair *i, j*, it is iterated over the length of $s_i$ + length of $s_j$, and then inside of that it is iterated for length of $s_j$. Since $c$ is the maximum length string in $S$, $2c \geq |s_i| + |s_j|$ and $c \geq |s_j|$. This means that for each pair the amount of iteration for said pair can be described as potentially better than $O(c^2)$. It is that complexity if all strings are the same length. Therefore, the complexity of *MISTAKES_FOR_PAIRS* is at worst $O(n^2c^2)$.

---

**Algorithm Code 1:** *Implementation of MISTAKES_FOR_PAIRS*

---

```
for i ∈ {0,..., n − 1}
    for j ∈ {0,..., n − 1}
        if i = j
            continue
        end if
        m ← S_j
        lr ← S_i
        for ind ∈ {0,..., |m| + |lr| − 1}
            for t ∈ {0,..., |m|}
                x ← ind - |m| + 1 + t
                if x ≥ 0 and x < |lr|
                    if m_t ≠ |lr|
                        MISTAKES_{i, j, ind} append x
                    end if
                end if
            end for
        end for
    end for
end for
```

---

**Part 2: Calculating the shortest SCS for each triple**

The next part of the algorithm is to calculate the $SCS_K$ for each triple. It will fill in $LMR_{l, m, r}$ for each triple where $LMR_{l, m, r}$ store the length of the $SCS_K$ for *l, m, r* where *m* possibly has mistakes. Let this algorithm be called *SCSK_FOR_TRIPLES*, implemented in Algorithm Code 2. Define three different, meaning three different indices, strings from *S*, *l, m,* and, *r*. The following algorithm will find the $SCS_K$ for this triple assuming that *m* is the string that will have the mistakes in it, meaning that *l* and *r* have no mistakes in them by the problem definition. To calculate the $SCS_K$ for each triple, first iterate over the three distinct indices to get *l, m,* and *r*. Now iterate over a new variable *len* that describes the length of the $SCS_K$ that is trying to be built. *l* and *r* are going to be touching the left and right edges respectively in the length *len* $SCS_K$ that is attempted to be constructed. Since *l* and *r* are going to be on their respective edges of the current $SCS_K$, and they cannot have mistakes, their overlap, if they overlap, must have no mistakes. It is possible to check this quickly using $|MISTAKES_{i, j, k}|$ that was calculated in the previous portion of the algorithm. Now that *l* and *r* are fit on the left and right edges of the current $SCS_K$ that is being built for this triple, it is time to see where *m* can go. Do this by iterating over all possible starting positions for *m,* given that it must end before or equal to where *r* is ending for the current value of *len.* So now the problem is to find how many mistakes *m* makes with *l* and/or *r* in its current fixed starting position. Note: it is important to also consider cases where the string with mistakes would be either the first or last element in the overall $SCS_K$. For this, do the same algorithm as described here but when there is only *m* and *l* or only *m* and *r.* Other than that, consider four cases:

**Case 1**: *m* fits entirely within *l*. In other words the last character of *m* given its starting position cannot go past the last character of *l. l* can intersect *r.* See figure 1 and figure 2 for examples.

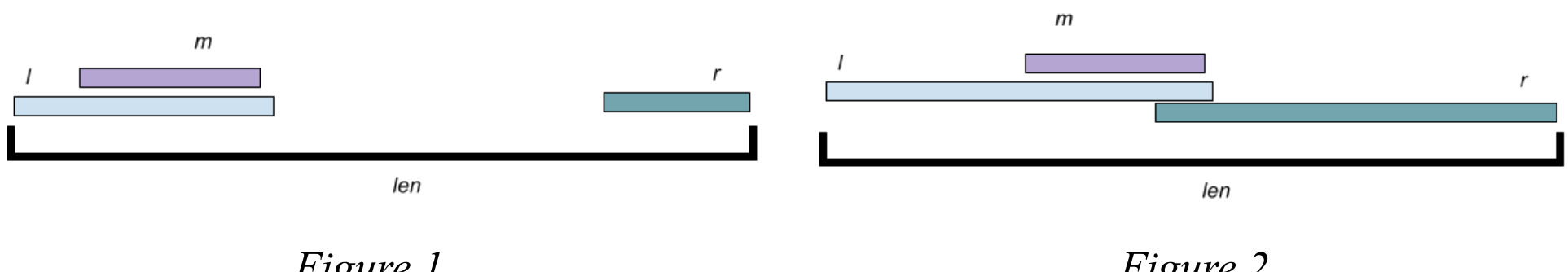

*Figure 1* *Figure 2*

In this case, simply use $|MISTAKES_{i,\ j,\ k}|$ that is already calculated for every pair to see how many mistakes *m* makes with *l* in the specific orientation. It does not matter if *r* intersects *l* as these two strings cannot have any mistakes, the intersecting part is the same, and therefore it is possible to use the $|MISTAKES_{i,\ j,\ k}|$ for the pair *l, m*. More detail will be provided in the implementation of this portion of the algorithm, in Algorithm Code 2.

**Case 2:** Similar to the last case, *m* fits entirely within *r.* So, the first character of *m* starts after or equal to the first character of *r*. *l* can intersect *r.* See figure 3 for an example.

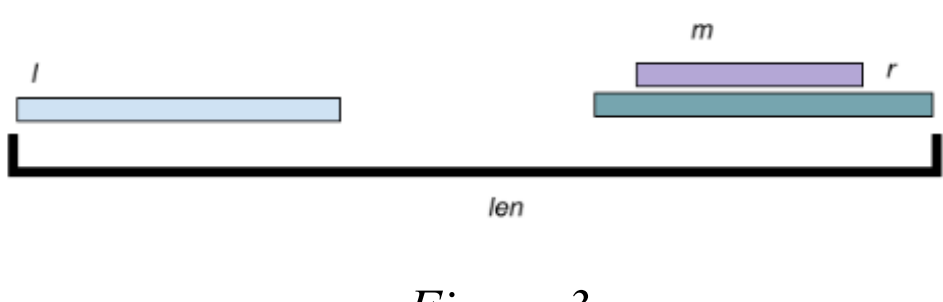

*Figure 3*

This case is almost entirely the same as Case 1. The only difference is that now the number of mistakes of *m* in the current position in *r* needs to be calculated. This is done using $|MISTAKES_{i,\ j,\ k}|$ once again, this time for *m* and *r*. Again, if *l* intersects *r* it does not matter for the same reason as described in Case 1.

**Case 3:** *l* and *r* do not intersect each other, *m* can intersect one, both, or neither. See figure 4 and figure 5 for an example.

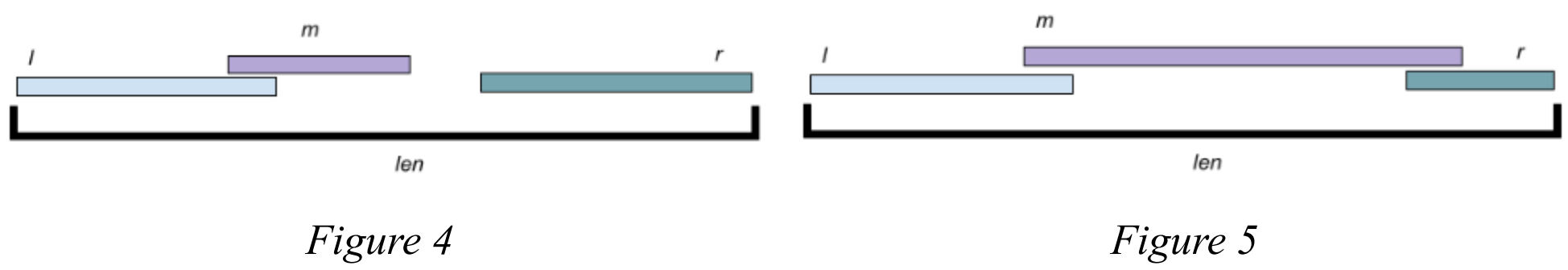

*Figure 4* *Figure 5*

This case is handled similarly to the previous two cases. Use $|MISTAKES_{i,\ j,\ k}|$ to get the mistakes in the *l, m* combination and the *r, m* combination. Then add those together.

**Case 4:** *l* and *r* intersect each other, and *m* intersects both *l* and *r.* See figure 6 for an example.

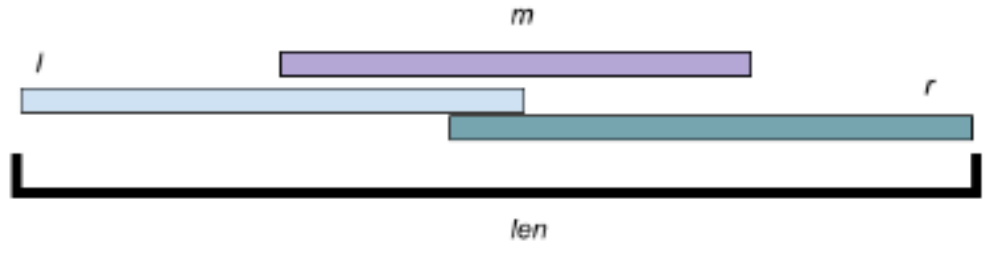

*Figure 6*

If $l$ and $r$ intersect each other, the situation becomes more complex than before. If the same approach as in Case 3 is used, mistakes in the region where $l, m, r$ intersect could be counted twice. To solve this problem the approach is to first do the same thing as in Case 3 but then subtract how many mistakes were counted twice. To figure this part out, it is key to remember that $MISTAKES_{i,\ j,\ k}$ is a list which stores the indices of the mistakes for $i, j$ given some position describing how the two strings are overlaid, $k$. So to find exactly how many mistakes need to be subtracted, perform a binary search on $MISTAKES_{r,\ m,\ k}$ to find how many indices in $m$ that have a mistake with $r$ are before or equal to the end of $l$. Note: due to how $MISTAKES_{i,\ j,\ k}$ is implemented, it is always a sorted list, thus binary search is applicable. More information on specific details can be seen in Algorithm Code 2.

**Lemma 2:** *Time complexity of SCSK_FOR_TRIPLES is $O(n^3c^2 logc)$ where c is the length of the longest string in S.*

*Proof:* First the algorithm iterates over the triple $l, m, r$. This is $O(n^3)$. Now in all cases, iterate over the total length of the $SCS_K$ that is trying to be built. This part can be said to work in $O(c)$ since the total length of the $SCS_K$ for three strings is at most three times the length of the longest string. Then iterate over the starting position of the middle string, $m$, and this can also be said to work in $O(c)$ for the same reason. Now look at cases. Cases 1 - 3 work in constant time since the only thing being done is accessing a list, which can be implemented as an array. Case 4 takes more time, as a binary search is required. It works in $O(logc)$ as there are at most $2c$ positions for each $MISTAKES_{i,\ j,\ k}$ pair. Thus, if worst case is assumed (Case 4 in all situations), then the complexity for the *SCSK_FOR_TRIPLES* algorithm is $O(n^3c^2 logc)$.

**Algorithm Code 2:** *Implementation of SCSK_FOR_TRIPLES*

```
for l ∈ {0,..., n − 1}
    for r ∈ {0,..., n − 1}
        for m ∈ {0,..., n − 1}
            if l = r or l = m or m = r
                continue
            end if
            L ← S_l
            M ← S_m
            R ← S_r
            for len ∈{ MAX(|L|, |R|),..., |L| + |R| + |M|}
                if len < |L| + |R| and |MISTAKES_{l, r, 2|L|+|R|- len - 1}|≠0
                    continue
                end if
                MS ← 0
                for start ∈ {0,..., len − |M|}
                    MS ← 0
                    ENDM ← start + |M| - 1
                    if start + |M| - 1 < |L|
                        MS ← MS + |MISTAKES_{l, m, ENDM}|
                    else if start ≥ len - |R|
                        ENDM ← ENDM - (len - |R|)
                        MS ← MS + |MISTAKES_{r, m, ENDM}|
                    else if len ≥ |L| + |R|
                        if start < |L|
                            MS ← MS +|MISTAKES_{l, m, ENDM}|
                        end if
                        ENDM ← ENDM - (len - |R|)
                        if ENDM ≥ 0
                            MS ← MS + |MISTAKES_{r, m, ENDM}|

                    else
                        MS ← MS +|MISTAKES_{l, m, ENDM}|
                        ENDM ← ENDM - (len - |R|)
                        MS ← MS + |MISTAKES_{r, m, ENDM}|
                        OVERLAP ← |L| + |R| - len
                        CT ← BINARYSEARCH(MISTAKES_{r, m, ENDM}, OVERLAP - 1)
                        MS ← MS - CT
                    end if
                    if MS ≤ K
                        LMR_{l, m, r} ← MIN(len, LMR_{l, m, r})
                    end if
                end for
            end for
        end for
    end for
end for
```

**Part 3: Computing standard SCS for each subset, keeping track of the first and last element**

The final part of the precalculation before the main algorithm is to calculate the normal *SCS* for each subset of *S*, for each string in *S* as the first and last element in the *SCS* for each subset. This part is based on the traveling salesman problem[11]. Call the algorithm for this precalculation *SCS_FIRST_LAST*. For example, the goal is to be able to tell what is the length of the normal *SCS* for the subset of *S* containing, for example, $s_0$, $s_3$, and, $s_7$ if $s_3$ is the last string(rightmost) in the *SCS*. But it is required to do this for all subsets and for every $s_i$ as the leftmost and rightmost of the subset *SCS*. To do this, a DP on bitmasks is needed. This version has some key modifications on the classical algorithm to align with the pre-calculated values from the previous sections and the main algorithm that is to come in Part 4. The main difference is that here it is not checked if one string is a substring of another - this detail will be handled later, in the main algorithm.

The first step, a precalculation, is to, for each pair of strings in *S*, calculate their maximum overlap with no mistakes - call this $OVERLAP_{w,\,v}$ where *w* is the left string and *v* is the right string in the pair. In other words, the overlap is between a suffix of *w* and a prefix of *v*. Then, let $DP_RIGHT_{i,\,j}$ store the length of the *SCS* for a subset of *S*, *i* that represents the subset through its bits, and string with index *j* be the rightmost in the *SCS*. The first step is to iterate over all number $i \in [1,\ 2^n)$. Then find all the bits set as 1 in *i*, and keep track of them as the indices of *S* in the current subset. More specifically, if *i* contains $2^j$, in its binary representation, then $s_j$ is in the current subset. The subset represents which strings in *S* the *SCS* is being calculated for. Once the indices in the subset are collected, iterate over all possible ending(rightmost) strings; they must be a part of the subset. Now, iterate over all second to last(second to rightmost) strings in the subset of *S*. Call the rightmost string *r* and the second to rightmost string from the subset *p*. Get the maximum overlap of *r* and *p* from $OVERLAP_{p,\,r}$. It is fine that it is not considered that *p* or *r* might be a substring of the other because these cases will be handled in the main algorithm, Part 4. So now to calculate $DP_RIGHT_{i,\,j}$ take the *DP_RIGHT* value of the subset without *r* and add on $|r|$ - $OVERLAP_{p,\,r}$. Do this process for all subsets, and since *i* is iterating in the increasing direction, all subsets that we would need to access for some other given subset will already be calculated. More details are in Algorithm Code 3 and 4. The *COUNT_OVERLAPS* method will be provided for reference in Algorithm Code 5.

**Algorithm Code 3:** *Implementation of FILL_DP_RIGHT*

```
for mask ∈ {1,..., 2^n − 1}
        for right ∈ {0,..., n − 1}
                if (right & mask ) = 0
                        continue
                end if
                for prevRight ∈ {0,..., n − 1}
                        if (prevRight & mask) = 0 or prevRight = right
                                continue
                        end if

                        DP_RIGHT_{mask, right} ← DP_RIGHT_{mask ⊕ 2^right} + |S_right| - OVERLAP_{prevRight, right}
                end for
        end for
end for
```

Also, let $DP_LEFT_{i,\ j}$ store the length of the *SCS* for a subset of *S*, *i* that represents the subset through its bits, and string with index *j* be the leftmost in the *SCS*. The algorithm is nearly exactly the same in this case as for $DP_RIGHT_{i,\ j}$, with now the leftmost instead of the rightmost element being iterated over. The implementation of *FILL_DP_LEFT* is very similar to that of *FILL_DP_RIGHT* and is provided in Algorithm Code 4.

**Algorithm Code 4:** *Implementation of FILL_DP_LEFT*

```
for mask ∈ {1,..., 2^n − 1}
        for left ∈ {0,..., n − 1}
                if (left & mask ) = 0
                        continue
                end if
                for prevLeft ∈ {0,..., n − 1}
                        if (prevLeft & mask) = 0 or prevLeft = left
                                continue
                        end if
                DP_LEFT_{mask, right} ← DP_LEFT_{mask ⊕ 2^left} + |S_left| - OVERLAP_{left, prevLeft}
                end for
        end for
end for
```

**Algorithm Code 5:** *Implementation of COUNT_OVERLAPS*

```
for w ∈ {0,..., n − 1}
        for v ∈ {0,..., n − 1}
                if v = w
                        OVERLAP_{w, v} ← |S_v|
                        continue
                end if
                for o ∈ {1 ,..., |S_w|}
                        if |MISTAKES_{v, w, o}| = 0
                                OVERLAP_{w, v} ← o
                        end if
                end for
        end for
end for
```

**Lemma 3:** *Time complexity of SCS_FIRST_LAST is* $O(n^2 2^n + n^2 c)$*. In other words*, *the time complexity of filling in* $DP_RIGHT_{i,j}$ , $DP_LEFT_{i,j}$*, and* $OVERLAP_{w,v}$ *is* $O(n^2(2^n + c))$*, where c is the length of the longest string in S.*

*Proof:* Start with the algorithm to fill in $OVERLAP_{w,v}$. Iterate over all pairs of strings in *S* in $O(n^2)$, then iterate over all possible overlaps in $O(c)$. For each overlap, quickly calculate if there is a mistake or not using $MISTAKES_{i,j,k}$ that was filled in previous sections. This part is $O(1)$. Now move on to $DP_RIGHT_{i,j}$ and $DP_LEFT_{i,j}$. The time complexities for them are the exact same, since the algorithm is the same, the only difference being whether or not to look at the rightmost elements or the leftmost elements in the subsets. In both cases, the first course of action is to iterate over all bitsets, which is done in $O(2^n)$. Then iterate over the rightmost and second rightmost or the leftmost and second leftmost. In both cases, this is worst case $O(n^2)$. Figuring out the bitmask that contains the subset without the leftmost or rightmost can be done in constant time through bitwise operations, so the overall time complexity to fill in $DP_RIGHT_{i,j}$ and $DP_LEFT_{i,j}$ is $O(n^2 2^n)$. Therefore, the time complexity for *SCS_FIRST_LAST* is

$O(n^2(2^n + c))$ as it is a combination of the time complexities proved here.

**Algorithm Code 6:** *Implementation of the SCS_FIRST_LAST algorithm*

---

*COUNT_OVERLAPS*
*FILL_DP_RIGHT*
*FILL_DP_LEFT*

---

**Part 4: Main algorithm; Getting the answer by iterating over triples and bitsets, using precomputed values**

All that is left is to utilize all of the precalculation in the main algorithm, call it *SCSK*. The idea of the algorithm goes as follows. First iterate over all triples in *S*. Call their indices *l, m, r* such that all three are distinct. These three will form our 'core' triple, with $s_m$ being the only string allowed to have mistakes. An important fact here is that if *l, m, r* is combined into their $SCS_K$, $LMR_{l, m, r}$, as described in previous sections, the $SCS_K$ problem for all *n* strings can be reduced down to a normal SCS problem. See Lemma 4:

**Lemma 4:** *If the three strings in S: $s_l$, $s_m$, $s_r$ are combined into their $SCS_K$-call it M, with $s_m$ being the string with mistakes, then the full $SCS_K$ problem, if $s_m$ is the string with mistakes, can be reduced down to a standard SCS problem with the strings in S minus $s_l$, $s_m$, $s_r$ plus the string M. More formally, if $s_m$ is the string with mistakes and $s_l$ is to the left of $s_m$ and $s_r$ is to the right of $s_m$, the full $SCS_K$ for the set S can be reduced down to a SCS problem with the set P:* $(S - \{s_l, s_m, s_r\}) \cup \{M\}$.

*Proof:* Since in this problem there is the constraint that no string in *S* is a substring of another in *S*, then no other string in *S* can engulf either *l* or *r.* Since *l* and *r* have no mistakes in them, then all strings placed on the outside of the core triple of *l, m, r* can simply be treated as they normally are in the *SCS* problem. There is a wondering of what happens if a string in *S* that is not *l, m, r* can be placed inside of the triple. Well, it cannot be a substring of *l, m, r* via problem description so the only option is it must be in between them but not full within any one: like between *l* and *m* or *m* and r. But if this was the optimal triple, it will be covered later when the triple of *l, m,* and the other string or *m, r,* and the other string are covered. This means that the core triple of *l, m, r*. Also, it is true that it is always optimal to only consider these core triples at their shortest $SCS_k$ position. In other words, if *l, m, r* can be combined with *k* mistakes in length X and Y, where X < Y, it is only needed to check the variation of length X for overall $SCS_K$. This is because since *l, r* have no mistakes and are arbitrarily spaced apart, and no string in *S* can completely overlap with *l* or *r* from problem constraints, putting *l, r* further apart will yield no benefit.

Then, iterate over all subsets of strings in *S* that remain after the previous removals. Call this subset *D*, and it is the subset of strings that will be to the left of the core triple. All remaining strings will be to the right of the core triple. Now to get the length of the overall $SCS_K$ use the length of the core triple and the values of *DP_RIGHT* and *DP_LEFT*. Then take the minimum length over all iterations. A detailed explanation is provided in the code, found in Algorithm Code 7. The general idea visualization can be seen in the diagram below. The time complexity of the *SCSK* method is $O(n^3 2^n)$, and is proved in lemma 5.

**Lemma 5:** *The time complexity of the SCSK method is* $O(n^3 2^n)$.

*Proof:* First, the algorithm iterates over all triples. This part is clearly $O(n^3)$. Then, the algorithm goes over all subsets. There are $2^n$ subsets for any set of size *n*, and if subsets are represented by integers in binary form as done here, the complexity of this part is $O(2^n)$. The procedure to find the length of the *SCSK* given the subset and the triple can be done in *O(1)* as only pre-calculated values are utilized. All operations that involve taking 2 to a power, $2^x$, can be done in *O(1)* through bitwise operations such as bitwise shifts, or could be quickly precalculated prior and stored in a list or array data structure. For more detail on implementation, see Algorithm Code 7 below.

---

**Algorithm Code 7:** *Implementation of the whole SCSK algorithm*

---

```
MISTAKES_FOR_PAIRS
SCSK_FOR_TRIPLES
SCS_FIRST_LAST

ANS ←NULL
for l ∈ {0,..., n − 1}
        for m ∈ {0,..., n − 1}
                for r ∈ {0,..., n − 1}
                        if l = r or l = m or m = r
                                continue
                        end if
                        for leftMask ∈ {0,..., 2^n − 1}
                                rightMask ← (2^n − 1) ⊕ leftMask
                                if (leftMask & 2^l) = 0 or (rightMask & 2^r) = 0
                                        continue
                                end if

                                length←LMR_{l, m, r} - |S_l| - |S_r| + DP_RIGHT_{leftMask, l} + DP_LEFT_{rightMask, r}
                                if ANS = NULL or length < ANS
                                        ANS ← length
                                end if
                        end for
                end for
        end for
end for
return ANS
```

---

**Theorem 1:** *Time complexity of the full SCSK algorithm is* $O(n^3 c^2 logc + n^3 2^n)$.

*Proof:* First, remember all of the time complexities of the previous sections. They are all independent, so the time complexities can be added together. Therefore, the overall time complexity of the *SCSK* algorithm as proposed is

$O(n^2 c^2 + n^3 c^2 logc + n^2 2^n + n^2 c + n^3 2^n)$, where $c$ is the length of the longest string in *S*. If only worst order complexities are kept, it can be simplified to $O(n^3 c^2 logc + n^3 2^n)$.